\documentclass[3p,times,twocolumn]{elsarticle}
 \biboptions{comma,sort&compress}
 
\usepackage{graphicx}
\usepackage{amsmath}
\usepackage{here}
\usepackage{ecrc}

\volume{00}

\firstpage{1}

\journalname{Nuclear and Particle Physics Proceedings}
\runauth{Giancarlo Rossi}

\jid{nppp}
\jnltitlelogo{Nuclear and Particle Physics Proceedings}

\usepackage{amssymb}

\usepackage[figuresright]{rotating}

\def\tr{\hbox{Tr}\,}

\newcommand{\beq}{\begin{equation}}
\newcommand{\eeq}{\end{equation}}
\newcommand{\beqn}{\begin{eqnarray}}
\newcommand{\eeqn}{\end{eqnarray}}
\newcommand{\nn}{\nonumber}
\def\slash#1{\mbox{$\not \!\!\! #1$}}
\def\Dslash{{\slash {\cal D}}}

\begin{document}

\begin{frontmatter}

\title{Elementary particle non-perturbative mass generation\\
A step towards a beyond-the-Standard-Model model} 

\cortext[cor0]{Talk presented at QCD22 - 25th International Conference in QCD (4-7/07/2022, Montpellier - France)}

 \author[label1]{Giancarlo Rossi}
 \address[label1]{Dipartimento di Fisica, Universit\`a di Roma Tor Vergata, Via della Ricerca Scientifica, 00133 Roma, Italy\\
INFN, Sezione di Roma Tor Vergata, Via della Ricerca Scientifica, 00133 Roma, Italy\\
Centro Fermi - Museo Storico della Fisica e Centro Studi e Ricerche Enrico Fermi, Via Panisperna 89a, 00184 Roma, Italy
}
\ead{rossig@roma2.infn.it}

\pagestyle{myheadings}
\markright{ }

\begin{abstract}
\noindent
We show that a recently discovered non-perturbative field-theoretical mechanism giving mass to elementary fermions, is also capable of generating a mass for the electro-weak bosons and can thus be used as a viable alternative to the Higgs scenario. A detailed analysis of this remarkable feature shows that the non-perturbatively generated fermion and $W$ masses have the parametric form $m_{f}\sim C_f(\alpha)\Lambda_{RGI}$ and $M_W\sim g_w c_w(\alpha)\Lambda_{RGI}$, respectively, where the coefficients $C_f(\alpha)$ and $c_w(\alpha)$ are functions of the gauge couplings, $g_w$ is the weak coupling and $\Lambda_{\rm RGI}$ is the RGI scale of the theory. In view of these expressions, we see that to match the experimental values of the top quark and $W$ masses, we are led to conjecture the existence of a yet unobserved sector of massive fermions (that we denote Tera-fermions) subjected, besides ordinary Standard Model interactions, to some kind of super-strong gauge interactions (Tera-interactions), so that the full theory (SM plus Tera-particles) will have an RGI scale $\Lambda_{\rm RGI}\equiv \Lambda_T$ in the TeV region. This approach offers a solution of the mass naturalness problem (there is no fundamental Higgs), an understanding of the fermion mass hierarchy and a physical interpretation of the electro-weak scale as a fraction of $\Lambda_T$.


\end{abstract}

\end{frontmatter}

\section{Introduction}
\label{sec:INTRO}

We have introduced in~\cite{Frezzotti:2014wja} and numerically explored in~\cite{Capitani:2019syo} a field-theoretical renormalizable model where an SU(2) fermion doublet, subjected to non-abelian gauge interactions of the QCD type, is coupled to a complex scalar field via a $d=4$ Yukawa term and an ``irrelevant'' $d>4$ Wilson-like operator. Despite the fact that both terms break chiral invariance, it was shown in~\cite{Frezzotti:2014wja} that there exists a critical value of the Yukawa coupling where chiral symmetry is recovered, up to small effects vanishing as the UV cut-off is removed. The interesting observation is that in the Nambu--Goldstone (NG) phase of this critical theory non-perturbative (NP) O($\Lambda_{\rm RGI}$) fermion masses get dynamically generated. 

They are a consequence of a sort of ``interference'' between UV chiral breaking terms and non-perturbative (NP) effects, coming from the spontaneous breaking of the (recovered) chiral symmetry occurring in the NG phase of the theory, that is standardly induced by the strong interaction dynamics. A detailed analysis of this remarkable field-theoretical feature shows that NP-ly generated fermion masses have the parametric form 
\begin{equation}
m_{f}\sim C_f(\alpha) \Lambda_{\rm RGI} \, ,\label{EQ1}
\end{equation}
where $\Lambda_{\rm RGI}$ is the RGI scale of the theory and $C_f(\alpha)$ is a function of the gauge coupling constants $\{\alpha\}$ of the theory. If we take the irrelevant chiral breaking Wilson-like term to be a $d=6$ operator, one finds to lowest loop order $C_f(\alpha) ={\mbox{O}}(\alpha_f^2)$, where $\alpha_f$ is the coupling constant of the strongest among the gauge interactions the particle is subjected to. 

An exact symmetry protects elementary particle masses against quantum power divergencies, at variance with what happens in Wilson lattice QCD (WLQCD). Masses are ``small'', i.e.\ ${\mbox{O}}(\Lambda_{\rm RGI})$, as the massless theory enjoys an enhanced (chiral) symmetry, in line with the 't Hooft naturalness notion~\cite{THOOFT}. 

A far reaching consequence of the formula~(\ref{EQ1}), if applied to the top quark, is that since $m_{top}={\mbox{O}}(\Lambda_{\rm RGI})$, to match its experimental values a new sector of super-strongly interacting particles, gauge-invariantly coupled to standard matter, needs to exist so that the complete theory, including the new sector and Standard Model (SM) particles, will have an RGI scale $\Lambda_{\rm RGI}\equiv\Lambda_{T}\gg\Lambda_{QCD}$ and of the order of a few TeV's. We immediately notice that the reason for assuming the existence of a super-strongly interacting sector here is very different from the reason why Technicolor was introduced~\cite{Susskind:1978ms,Weinberg:1979bn}. Technicolor was invoked to give mass to the EW bosons in the first place, while in the present approach super-strong interactions are introduced to give the right order of magnitude to the top quark and $W$ masses. Anyway to avoid confusion and following Glashow~\cite{Glashow:2005jy}, we will refer to this new set of particles as Tera-particles. 

The model can be naturally extended to incorporate EW interactions and leptons. EW bosons  as well as leptons will acquire NP masses O($\Lambda_{T}$) via the same mechanism that leads to eq.~(\ref{EQ1})~\footnote{With the SM hypercharge assignment, in our model neutrinos are massless because $\nu_R$ is sterile. We will not discuss here how to give mass to neutrinos in the present framework.}, but with coefficient functions that scale like powers of the EW gauge coupling, $g_w$. For instance, to the leading loop order one gets for the $W$ mass (see sect.~\ref{sec:WMASS} and refs.~\cite{Frezzotti:2018zsy,Frezzotti:2019npa})
\begin{equation}
M_W\sim \sqrt{\alpha_w} c_w(\alpha)\Lambda_{\rm RGI}\, ,\quad c_w(\alpha) = {\mbox{O}}(\alpha)\, ,
\label{EQ2}
\end{equation} 
where $\alpha$ in eq.~(\ref{EQ2}) is a short-hand for the set of gauge couplings $[\alpha_s,\alpha_T]$ with $\alpha_s$ and $\alpha_T$ the strong and Tera-strong gauge couplings, respectively. 

Eqs.~(\ref{EQ1}) and~(\ref{EQ2}) are somewhat similar to the expression of the Higgs-masses of fermions and $W$'s with, however, two fundamental differences. The first is that the scale of the mass is not the vev of the Higgs field, but a dynamical scale related to a new interaction. The second is that the modulation of the Yukawa couplings that in the SM is introduced by hand to accomodate the values of fermion masses, here, as we said before, is controlled to leading order by the magnitude of the gauge coupling of the strongest among the interactions the particle feels. 

We conclude from this analysis that the NP scenario for mass generation we are advocating can be considered as a valid alternative to the Higgs mechanism, with the extra advantage that we will not have to deal with power divergencies in the Higgs mass as there is no Higgs around. A further conceptual bonus is that, in view of the above mass formulae, we have a natural interpretation of the magnitude of the EW scale, as (a fraction of) the dynamical physical parameter, $\Lambda_T$.

There is a number of further interesting features of the approach we are describing that are worth mentioning, but which we are not going to develop in this contribution for lack of space.

First of all, the dependence of the NP fermion masses upon the gauge couplings (to leading order one finds $C(\alpha_f) \!=\!{\mbox{O}}(\alpha_f^2)$ for a $d=6$ Wilson-like operator, see eq.~(\ref{LWIL}) below) may offer a hint to understand the fermion mass hierarchy $m_\tau\! \ll \!m_t \!\ll \!m_T$, as due to the ranking among weak, strong and Tera-strong gauge couplings, $\alpha_Y\! \ll\! \alpha_s \!\ll \!\alpha_T$. 
We expect the gauge coupling dependence of the NP fermion mass estimate~(\ref{EQ1}) to be more compelling for the heaviest of the SM fermion families which we assume the following analysis refers to. 

Secondly, lacking the need for a Higgs boson, we propose to interpret the 125~GeV resonance, recently identified at LHC, as a $W^+W^-/ZZ$ composite state, bound by exchanges of Tera-particles. Since this bound state is light on the $\Lambda_{T}$ scale, it should be incorporated in the low energy effective Lagrangian (LEEL) that one gets by integrating out the ``heavy'' Tera-degrees of freedom. As a result, at (momenta)$^2 \ll \Lambda_T^2$ the effective Lagrangian is seen to resemble very much to the Lagrangian of the SM (see sect.~\ref{sec:ITDF} and ref.~\cite{NEXT}). 

Finally it was shown in ref.~\cite{Frezzotti:2016bes} that with a reasonable choice of the elementary particle content, a theory extending the SM with the inclusion of the new super-strong sector leads to gauge coupling unification at a $\sim 10^{18}$~GeV scale. This higher than usual unification scale, irrespective of the underlying GUT theory generically allowing for proton decay, is likely to yield a proton life-time comfortably larger than the present limit $\tau_{\rm prot} > 1.7 \times10^{34}$ years.

It goes without saying that the issue of how eq.~(\ref{EQ1}) could be extended to deal with lighter families (and to cope with weak isospin splitting) is outside the scope of this talk and it is still a matter of investigation. 

\vspace{-.2cm}
\section{A simple toy-model}
\label{sec:STM}

The simplest model enjoying the NP mass generation mechanism we outlined before is described by a Lagrangian where an SU(2) fermion doublet, subjected to non-abelian gauge interactions (of the QCD type), is coupled to a complex scalar via $d=4$ Yukawa and ``irrelevant'' $d=6$ Wilson-like chiral breaking terms. The Lagrangian of this toy-model reads 
{\small \begin{eqnarray}
\hspace{-1.cm}&&{\cal L}_{\rm toy}(q,A;\Phi)=\nn \\
\hspace{-1.cm}&&\quad={\cal L}_{kin}(q,A;\Phi)\!+\!{\cal V}(\Phi)
\!+ \!{\cal L}_{Yuk}(q;\Phi)\!+\!{\cal L}_{Wil}(q,A;\Phi) \label{SULL}\\
\hspace{-1.cm}&&\bullet\,{\cal L}_{kin}(q,A;\Phi)= \frac{1}{4}(F^A\cdot F^A)+\bar q_L \,\Dslash^A q_L+\bar q_R\, \Dslash^A q_R+\nn\\
\hspace{-1.cm}&&\quad+\frac{1}{2}{\tr}\big{[}\partial_\mu\Phi^\dagger\partial_\mu\Phi\big{]}\label{LKIN}\\
\hspace{-1.cm}&&\bullet\,{\cal V}(\Phi)= \frac{\mu_0^2}{2}{\tr}\big{[}\Phi^\dagger\Phi\big{]}+\frac{\lambda_0}{4}\big{(}{\tr}\big{[}\Phi^\dagger\Phi\big{]}\big{)}^2\label{VPHI}\\
\hspace{-1.cm}&&\bullet\,{\cal L}_{Yuk}(q;\Phi)= \eta\,\big{(} \bar q_L\Phi q_R+\bar q_R \Phi^\dagger q_L\big{)} \label{LYUK}\\
\hspace{-1.cm}&&\bullet\,{\cal L}_{Wil}(q,A;\Phi)\!=\! \frac{b^2}{2}\rho\,\big{(}\bar q_L\overleftarrow{\cal D}^A_\mu\Phi {\cal D}^A_\mu q_R\!+\!\bar q_R \overleftarrow{\cal D}^A_\mu \Phi^\dagger {\cal D}^A_\mu q_L\big{)}\label{LWIL}\, ,
\end{eqnarray}
}
where $\Phi \!=\! \varphi_01\!\!1\!+\!i \varphi_j\tau^j\!=\! [-i\tau_2 \varphi^\star| \,\varphi]$ is a $2\!\times \!2$ matrix with $\varphi\!=\!(\varphi_2\!-\!i\varphi_1,\varphi_0\!-\!i\varphi_3)^T$ a complex scalar doublet, singlet under the color gauge, SU($N_c$), $b^{-1} \!\sim\! \Lambda_{UV}$ is the UV cutoff, $\eta$ is the Yukawa coupling, $\rho$ is to keep track of ${\cal L}_{Wil}$ and ${\cal D}_\mu^A$ the covariant derivative. The Lagrangian~(\ref{SULL}) is power counting renormalizable (like WLQCD is, despite the presence of the $d\!=\! 5$ Wilson term). 

Among other obvious symmetries, ${\cal L}_{\rm toy}$ is invariant under the 
(global) transformations involving fermions and scalars ($\Omega_{L/R} \in {\mbox{SU}}(2)$)
\begin{eqnarray}
\hspace{-1.3cm}&&\chi_L\!\times\! \chi_R \!= \! [\tilde\chi_L\times (\Phi\to\Omega_L\Phi)]\times [\tilde\chi_R\times (\Phi\to\Phi\Omega_R^\dagger)] \, ,
\label{CHIL}\\
\hspace{-1.3cm}&&\tilde\chi_{L/R} : q_{L/R}\rightarrow\Omega_{L/R} q_{L/R} \, ,\quad \bar q_{L/R}\rightarrow \bar q_{L/R}\Omega_{L/R}^\dagger \, . \label{GTWT}
\end{eqnarray}
The exact $\chi_L\times \chi_R $ symmetry can be realized either {\it \'a la} Wigner or {\it \'a la} NG depending on the shape of the scalar potential. In any case no power divergent fermion mass can be generated by quantum corrections as the mass operator $(\bar q_L q_R+ \bar q_R q_L)$ is not invariant under $\chi_L\times\chi_R$. 

The operators ${\cal L}_{Wil}$ and ${\cal L}_{Yuk}$ break $\tilde\chi_{L} \times \tilde\chi_{R}$ and mix. Thus for generic values of $\eta$, ${\cal L}_{\rm toy}$ is not invariant under the fermionic chiral transformations $\tilde\chi_{L} \times \tilde\chi_{R}$. However, one can show~\cite{Frezzotti:2013raa,Frezzotti:2014wja} that chiral invariance can be recovered (up to vanishingly small O($b^{2}$) terms) at a critical value $\eta=\eta_{cr}$, where the Wilson-like term and the Yukawa terms ``compensate'', much like chiral symmetry is recovered (up to O($a$) cutoff effects) in WLQCD by tuning the bare quark mass to a critical value, $m_{cr}$, where the Wilson and the mass term ``compensate''. The condition determining $\eta_{cr}$ corresponds to the vanishing of the effective Yukawa term in the Quantum Effective Lagrangian (QEL) of the theory~\footnote{By Quantum Effective Lagrangian we mean the generating functional of the 1PI vertices of the theory.}. 

\begin{figure}[htbp]    
\centerline{\includegraphics[scale=0.35,angle=0]{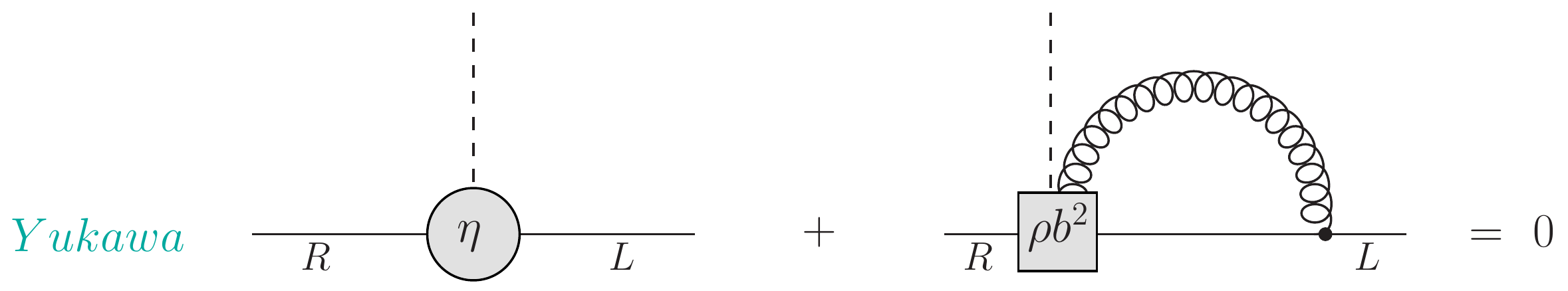}}
\centerline{\includegraphics[scale=0.35,angle=0]{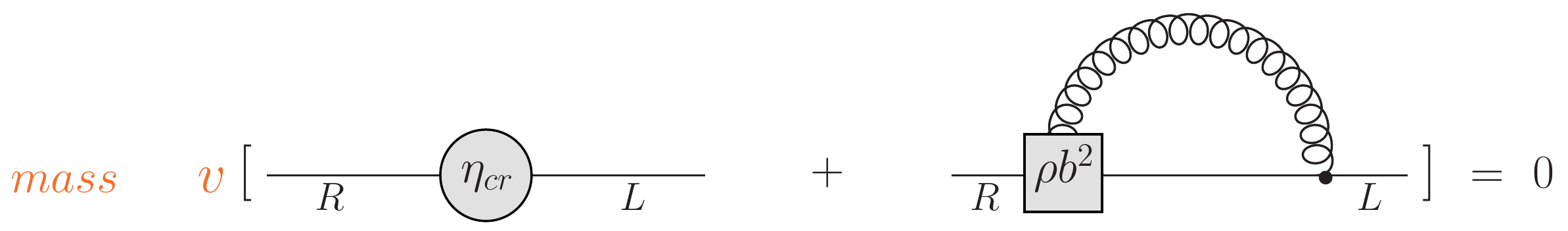}}
\caption{\small{Top: the cancellation of the effective Yukawa term implied by the condition determining $\eta_{cr}$ in the Wigner phase. The disc represents the Yukawa vertex, the box the insertion of the Wilson-like vertex. Bottom: the mechanism behind the cancellation of the Higgs-like mass of the quark in the NG phase occurring at  the $\eta=\eta_{cr}$.}}
\label{fig:fig1}
\end{figure}

In the Wigner phase (where $\langle |\Phi|^2\rangle=0$) this condition implies the cancellation diagrammatically represented at the lowest order in the top panel of fig.~\ref{fig:fig1} with complete decoupling of fermions and scalars~\cite{Frezzotti:2014wja}.

In the NG phase (where $\langle |\Phi|^2\rangle\!=\!v^2\!=\!\mu_0^2/\lambda_0$) the same condition implies that the Higgs-like mass of the fermion is cancelled out, as schematically shown in the bottom panel of fig~\ref{fig:fig1}. It is a peculiar feature of the model~(\ref{SULL}) that the (quadratic) divergencies in the loop integrals in fig.~\ref{fig:fig1} are exactly compensated by the $b^2$ factors coming from the insertion of the Wilson-like vertex. In the following we will encounter other instances of this UV vs.\ IR compensation.

\subsection{NP mass generation}
\label{sec:NPMG}

Now the key observation is that at $\eta=\eta_{cr}$ where  in the NG phase the fermion Higgs-mass is cancelled, the quark acquires a non-vanishing NP mass via a UV vs.\ IR compensation mechanism,  reminiscent of the one that makes finite the 1-loop diagrams of fig.~\ref{fig:fig1}. 

Because of this kind of subtle UV vs.\ IR compensations, it is crucial to analyze and determine the structure of the formally O($b^2$) terms of the regularized theory. This can be properly done by making recourse to the Symanzik expansion technique~\cite{SYM}. 

\begin{figure}[htbp]    
\centerline{\includegraphics[scale=0.35,angle=0]{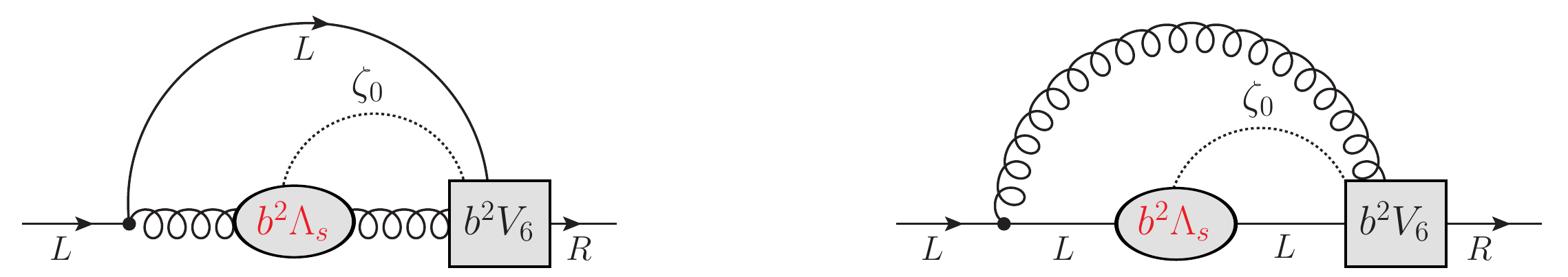}} 
\caption{\small{Two lowest loop order quark self-energy diagrams. The blobs represent insertions of $\Delta{\cal L}_{NP}^{Sym}$}.}
 \label{fig:fig3}
\end{figure}

This analysis was carried out in ref.~\cite{Frezzotti:2014wja} where it was shown that, in order to saturate the $\tilde\chi_L\times\tilde\chi_R$ Ward--Takahashi Identities in the presence of the spontaneous breaking of the (restored) $\tilde\chi_L\times\tilde\chi_R$ symmetry, one must include in the Symanzik expansion also formally O($b^2$) NP-ly generated operators of the form 
\beqn
&&O_{6,\bar q q} \propto b^2 \Lambda_s \alpha_s |\Phi|\, \Big{[}\bar q\,\, \Dslash^{A} q\Big{]} \, ,\label{SOP1}\\
&&O_{6,FF} \propto b^2\Lambda_s\alpha_s |\Phi| \,\Big{[} F^A\!\cdot\!F^A \Big{]} \, .\label{SOP2}
\eeqn
The expression of these operators is completely fixed by symmetries (in particular $\chi_L\times\chi_R$) and dimensional considerations. The presence of the $\Lambda_s$ factor signals their NP origin. The existence of the operators~(\ref{SOP1}) and~(\ref{SOP2}) can be effectively taken into account by including in the formalism the new diagrams generated by the augmented Lagrangian 
\beqn
\hspace{-1.4cm}&&{\cal L}_{\rm toy} \to {\cal L}_{\rm toy} + \Delta{\cal L}_{NP}^{Sym} \, , \nn\\
\hspace{-1.4cm}&&\Delta{\cal L}_{NP}^{Sym} = b^2\Lambda_s\alpha_s |\Phi|\,\Big{[}c_{FF} F^A\!\cdot\! F^A+c_{\bar q q}\bar q\,\, \Dslash^{A} q\Big{]}+\ldots \label{LAUG}
\eeqn
The remarkable fact is that new diagrams are generated that contribute, among other correlators, to the quark self-energy. A couple of them are drawn in fig.~\ref{fig:fig3}. A straightforward calculation gives for the amputated zero momentum diagram in the right panel of fig.~\ref{fig:fig3}
\begin{eqnarray}
\hspace{-1.3cm}&&{{m_q^{NP}}}\!\propto\! {{\alpha_s^2}} \! \int^{1/b}\!\frac{d^4 k}{k^2}\frac{\gamma_\mu k_\mu}{k^2}\int^{1/b}\!\!\!\frac{d^4 \ell}{\ell^2+m^2_{\zeta_0}} \frac{\gamma_\nu(k+\ell)_\nu}{(k+\ell)^2} \cdot 
 \nn\\
\hspace{-1.3cm}&&\cdot \,b^2\gamma_\rho (k+\ell)_\rho \,b^2{{\Lambda_{s}}}\gamma_\lambda (2k+\ell)_\lambda \sim {{\alpha_s^2 \Lambda_{s}}}\, .\label{MQ}
\end{eqnarray}
Here again we get a finite result, owing to an exact compensation between the UV power divergencies of the 2-loop integrals and the IR behaviour determined by the expression of the Wilson-like vertex~(\ref{LWIL}) and the physics of the spontaneous breaking of the chiral $\tilde\chi_{L} \times \tilde\chi_{R}$ symmetry, in turn encoded in the form of the Symanzik operators~(\ref{SOP1}) and~(\ref{SOP2}). A similar compensation occurs in the calculation of the diagram in the left panel of fig.~\ref{fig:fig3}.

\subsection{The QEL of the critical theory in the NG phase}
\label{sec:QNGC}

The resulting non-vanishing quark mass, the occurrence of which was checked in the explicit lattice simulations of the model~(\ref{SULL}) in ref.~\cite{Capitani:2019syo}, can be incorporated in the QEL, $\Gamma^{NG}$, of the theory, upon introducing the non-analytic field ${U}$ in the polar decomposition 
\beq
\Phi = (v+\zeta_0)U\, , \quad U=\exp[i\vec\tau\vec \zeta/v]\, .\label{PHPH}
\eeq
Simply on the basis of symmetries and observing that $U$ transforms like $\Phi$, so that new $\chi_L\times\chi_R$ invariant operators can be constructed~\cite{Frezzotti:2014wja,NEXT}, one obtains 
\beqn
\hspace{-1.2cm}&& \Gamma_{d=4}^{NG}\! =\! \Gamma_{d=4}^{Wig}\Big{|}_{\hat \mu_\Phi^2 < 0} +
\frac{1}{4}(F^A\!\cdot\! F^A)\!+\!\bar q_L\, \Dslash^{A} q_L\!+\!\bar q_R\, \Dslash^{A} q_R +\nn\\
\hspace{-1.2cm}&& +{{c_q \Lambda_s [\bar q_L U q_R\! +\! \bar q_R U^\dagger q_L]}} \!+\! \frac{c^2\Lambda_s^2}{2} \tr[\partial_\mu U^\dagger\partial_\mu U] \, . \label{GW}
\eeqn
Naturally, of special interest is the fourth term in the r.h.s.\ of eq.~(\ref{GW}) because, upon expanding $U =\!1\!\!1\!+ \vec \tau\,\vec\zeta/v+\ldots$, it gives rise to a mass for the fermion plus a wealth of NG boson interactions.

\section{Introducing weak and Tera-interactions}
\label{sec:WMASS}

To proceed to the construction of a possible, realistic beyond-the-SM-model we clearly need to introduce EW interactions. At the same time, as mentioned in the Introduction, it is also necessary to extend the model by incorporating a super-strongly interacting sector, in order for the whole theory to have an RGI scale, $\Lambda_{T}$, much larger than $\Lambda_{QCD}$ and of the order of a few TeVs. Only in this way there is the chance that eq.~(\ref{EQ1}) can yield the correct order of magnitude of the top quark mass. As proved in~\cite{GCRNEW}, the extended Lagrangian is obtained by doubling the structure of quarks in order to encompass Tera-particles ($Q$=Tera-quarks and $G$=Tera-gluons) and gauging the exact $\chi_L$ symmetry to introduce weak interactions. The whole Lagrangian then reads 
{\small{\beqn
\hspace{-1.4cm}&&{\cal L}(q,Q;\Phi;A,G,W)={\cal L}_{kin}(q,Q;\Phi;A,G,W)+{\cal V}(\Phi)+\nn\\
\hspace{-1.4cm}&&\quad+{\cal L}_{Yuk}(q,Q;\Phi) +{\cal L}_{Wil}(q,Q;\Phi;A,G,W) \label{SULLWQ}\\
\hspace{-1.4cm}&&\bullet\,{\cal L}_{kin}
= \frac{1}{4}\Big{(}F^A\cdot F^A+F^G\cdot F^G+F^W\cdot F^W\Big{)}+\nn\\
\hspace{-1.4cm}&&\quad+\Big{[}\bar q_L\,/\!\!\!\!{\cal D}^{AW} q_L\!+\!\bar q_R\,/\!\!\!\!{\cal D}^{A} q_R\Big{]}\!+\!\Big{[}\bar Q_L\,/\!\!\!\!{\cal D}^{AGW} Q_L\!+\!\bar Q_R\,/\!\!\!\!{\cal D}^{AG} Q_R\Big{]}+\nn\\
\hspace{-1.4cm}&&\quad+\frac{k_b}{2}{\tr}\big{[}({\cal D}\,^{W}_\mu \Phi)^\dagger{\cal D}^W_\mu\Phi\big{]}\label{LKINWQ}\\
\hspace{-1.4cm}&&\bullet\,{\cal V}
= \frac{\mu_0^2}{2}k_b{\tr}\big{[}\Phi^\dagger\Phi\big{]}+\frac{\lambda_0}{4}\big{(}k_b{\tr}\big{[}\Phi^\dagger\Phi\big{]}\big{)}^2\label{VWQ}\\
\hspace{-1.4cm}&&\bullet\,{\cal L}_{Yuk}
=\eta_q\,\big{(} \bar q_L\Phi\, q_R\!+\!\bar q_R \Phi^\dagger q_L\big{)} \!+ \!\eta_Q\,\big{(} \bar Q_L\Phi\, Q_R\!+\!\bar Q_R \Phi^\dagger Q_L\big{)}\label{LYUKWQ} \\
\hspace{-1.4cm}&&\bullet\,{\cal L}_{Wil}
\!=\! \frac{b^2}{2}{\rho_q} \big{(}\bar q_L{\overleftarrow {\cal D}}\,^{AW}_\mu\Phi {\cal D}^A_\mu q_R\!+\!\bar q_R \overleftarrow{\cal D}\,^A_\mu \Phi^\dagger {\cal D}^{AW}_\mu q_L\big{)}\!+\!\nn\\
\hspace{-1.4cm}&&\quad+\frac{b^2}{2}{\rho_Q}\,\big{(}\bar Q_L{\overleftarrow {\cal D}}\,^{AGW}_\mu\Phi {\cal D}^{AG}_\mu Q_R+\bar Q_R \overleftarrow{\cal D}\,^{AG}_\mu \Phi^\dagger {\cal D}^{AGW}_\mu Q_L\big{)}\label{LWILWQ}
\eeqn}}
with obvious notations for the covariant derivatives. Besides the Yukawa (eq.~(\ref{LYUKWQ})) and the Wilson-like~(eq.~(\ref{LWILWQ})) operators, now also the kinetic term of the scalar (last term in eq.~(\ref{LKINWQ})) breaks $\tilde\chi_L\times \tilde\chi_R$ and mixes with ${\cal L}_{Yuk}$ and ${\cal L}_{Wil}$. Thus to get to the critical theory (invariant under $\tilde\chi_L\times \tilde\chi_R$), on top of $\eta_q$ and $\eta_Q$, a further parameter, $k_{b}$, needs to be introduced and appropriately tuned. The conditions determining the critical theory correspond to have vanishing effective Yukawa interactions  and vanishing scalar kinetic term in the QEL. Notice that the coefficient $k_b$ appears also in the expression of the scalar potential~(\ref{VWQ}). The reason for this is that in this way the (bare) vev$^2$ (in the NG phase) and the quartic coupling of the canonically normalized scalar field will have the standard definitions, $v^2=|\mu^2_0|/\lambda_0$ and $\lambda_0$.

\begin{figure}[htbp]
\centerline{\includegraphics[scale=0.3,angle=0]{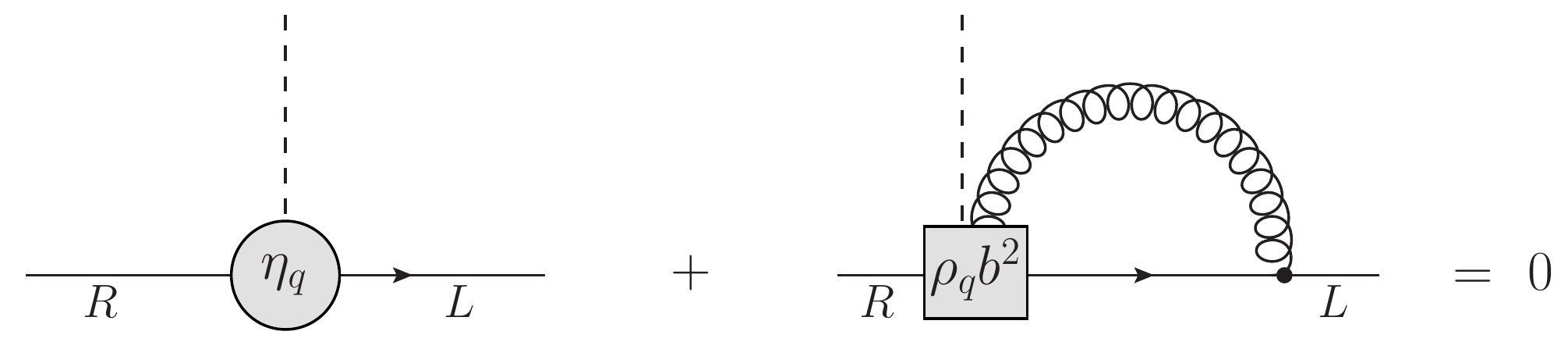}}\centerline{\includegraphics[scale=0.3,angle=0]{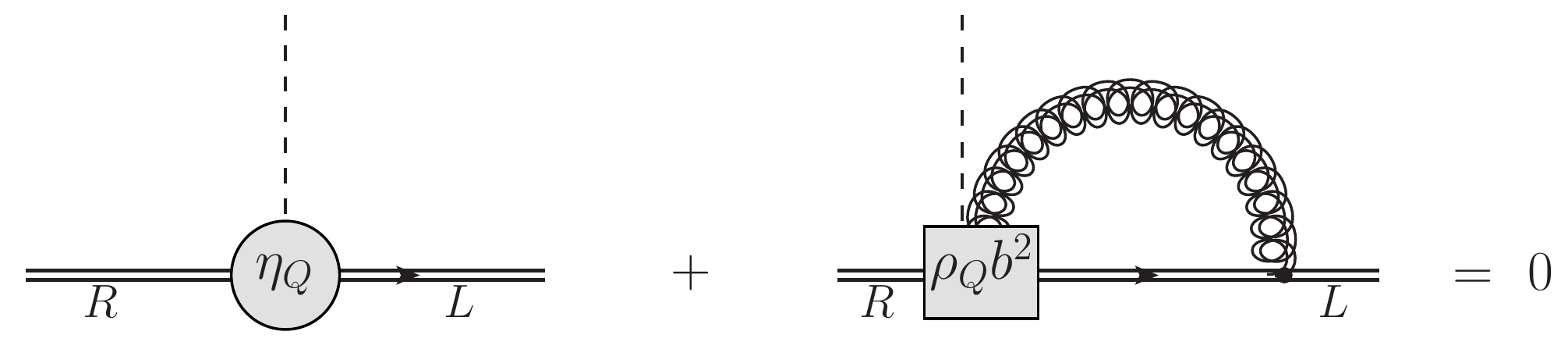}}
\vspace{.2cm}
\centerline{\includegraphics[scale=0.28,angle=0]{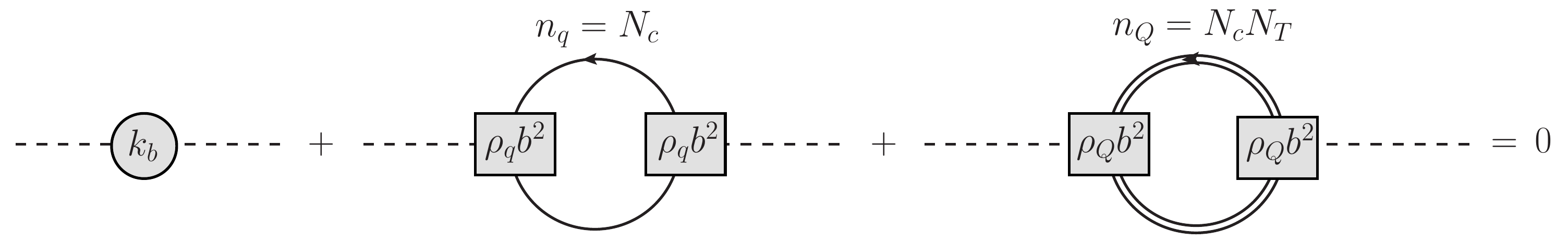}}
\caption{\small{From top to bottom, the cancellation implied by the tuning conditions determining $\eta_{q\,cr}$, $\eta_{Q\,cr}$ and  $k_{b\,cr}$ at the lowest order in the Wigner phase. Boxes represent the insertion of the quark and Tera-quark Wilson-like vertices, respectively, the disc  the insertion of the scalar kinetic term. Double lines represent Tera-particles. The rest of the notations is as in fig.~\ref{fig:fig1}.}}
\label{fig:fig4a}
\end{figure}

The tuning conditions that determine $\eta_{q\,cr}$, $\eta_{Q\,cr}$ and $k_{b\,cr}$ in the Wigner phase imply the cancellations schematically illustrated in the fig.~\ref{fig:fig4a}. To the lowest loop order one finds
{\small \beqn
\hspace{-1.cm}&&\eta_{q\,cr}^{(1-loop)}\!=\!\rho_q \, \eta_{1q} \alpha_s \, , \qquad \eta_{Q\,cr}^{(1-loop)}\!=\!\rho_Q \, \eta_{1Q} \alpha_T \, ,\label{ETAT1}\\
\hspace{-1.cm}&&\hspace{1.3cm}k_{b\,cr}^{(1-loop)}\!=\! [\rho_q^2 N_c+ \rho_Q^2 N_c  N_T]k_1 \, ,
\eeqn}
with $\eta_{1q}$,  $\eta_{1Q}$ and $k_1$ computable coefficients and SU($N_T$) the Tera-gauge group. As before, UV loop divergencies are exactly compensated by the IR behaviour of the inserted Wilson-like vertices. 

\vspace{-.2cm}

\begin{figure}[htbp]
\centerline{\includegraphics[scale=0.35,angle=0]{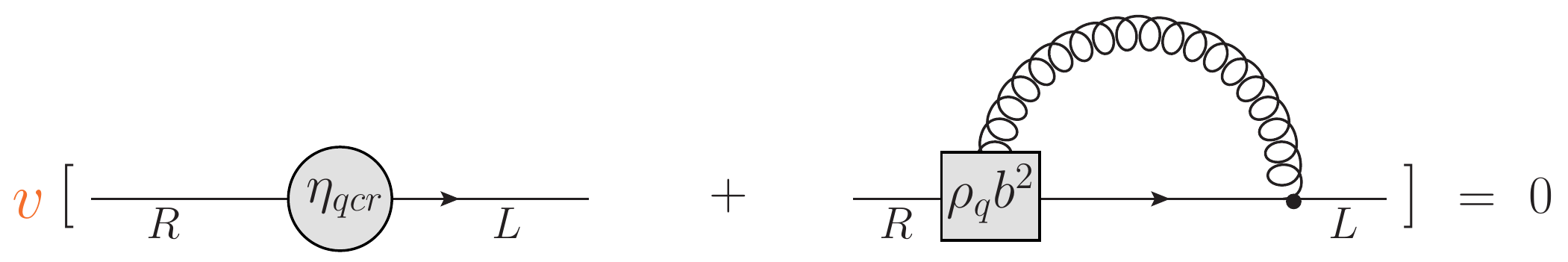}}\centerline{\includegraphics[scale=0.35,angle=0]{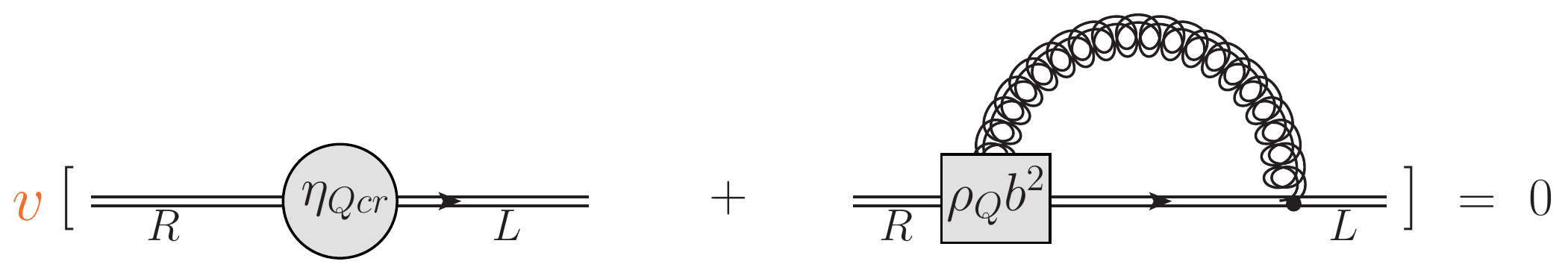}}
\centerline{\includegraphics[scale=0.25,angle=0]{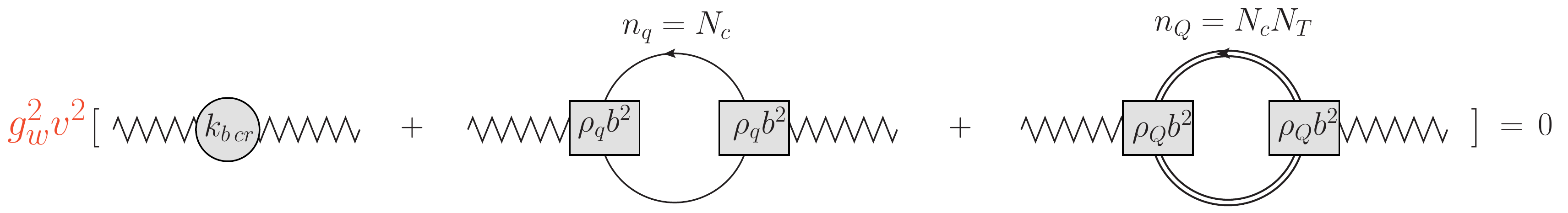}}
\caption{\small{From top to bottom, the cancellation mechanism of the Higgs-like mass of quark, Tera-quark and $W$ occurring in the NG phase of the critical theory.
}}
\label{fig:fig5a}
\end{figure}

In the NG phase the tuning conditions entail the cancellation of the Higgs-like mass of quarks, Tera-quarks and $W$'s, as schematically represented in figs.~\ref{fig:fig5a}. 

Extending to the present case the Symanzik analysis of the formally O($b^2$) operators necessary to describe the NP effects related to the spontaneous breaking of the (recovered) $\tilde\chi_L\times\tilde\chi_R$ symmetry in the regularized theory, one finds the following set of operators
{\small \beqn
\hspace{-1.4cm}&&O_{6,\bar Q Q}^s = b^2 \alpha_s\,\rho_Q\Lambda_T|\Phi| \Big{[}\bar Q_L\,\Dslash^{AGW} Q_L+\bar Q_R\,\Dslash^{AG} Q_R\Big{]} \label{OTT}\\
\hspace{-1.4cm}&&O_{6,\bar Q Q}^T = b^2 \alpha_T\,\rho_Q\Lambda_T|\Phi| \Big{[}\bar Q_L\,\Dslash^{AGW} Q_L+\bar Q_R\,\Dslash^{AG} Q_R\Big{]} \label{OSS} \\
\hspace{-1.4cm}&&O_{6,AA} = b^2\alpha_s\, \rho_Q\Lambda_T|\Phi| F^A\!\cdot\!F^A\label{OAA}\\
\hspace{-1.4cm}&&O_{6,GG} = b^2 \alpha_T\,\rho_Q\Lambda_T |\Phi| F^G\!\cdot\!F^G \label{OGG} \\
\hspace{-1.4cm}&&O_{6,WW} = b^2\alpha_w\,\rho_Q\Lambda_T|\Phi| F^W\!\cdot\!F^W\label{OWW}
\eeqn}
\begin{figure}[htbp]
\centerline{\includegraphics[scale=0.30,angle=0]{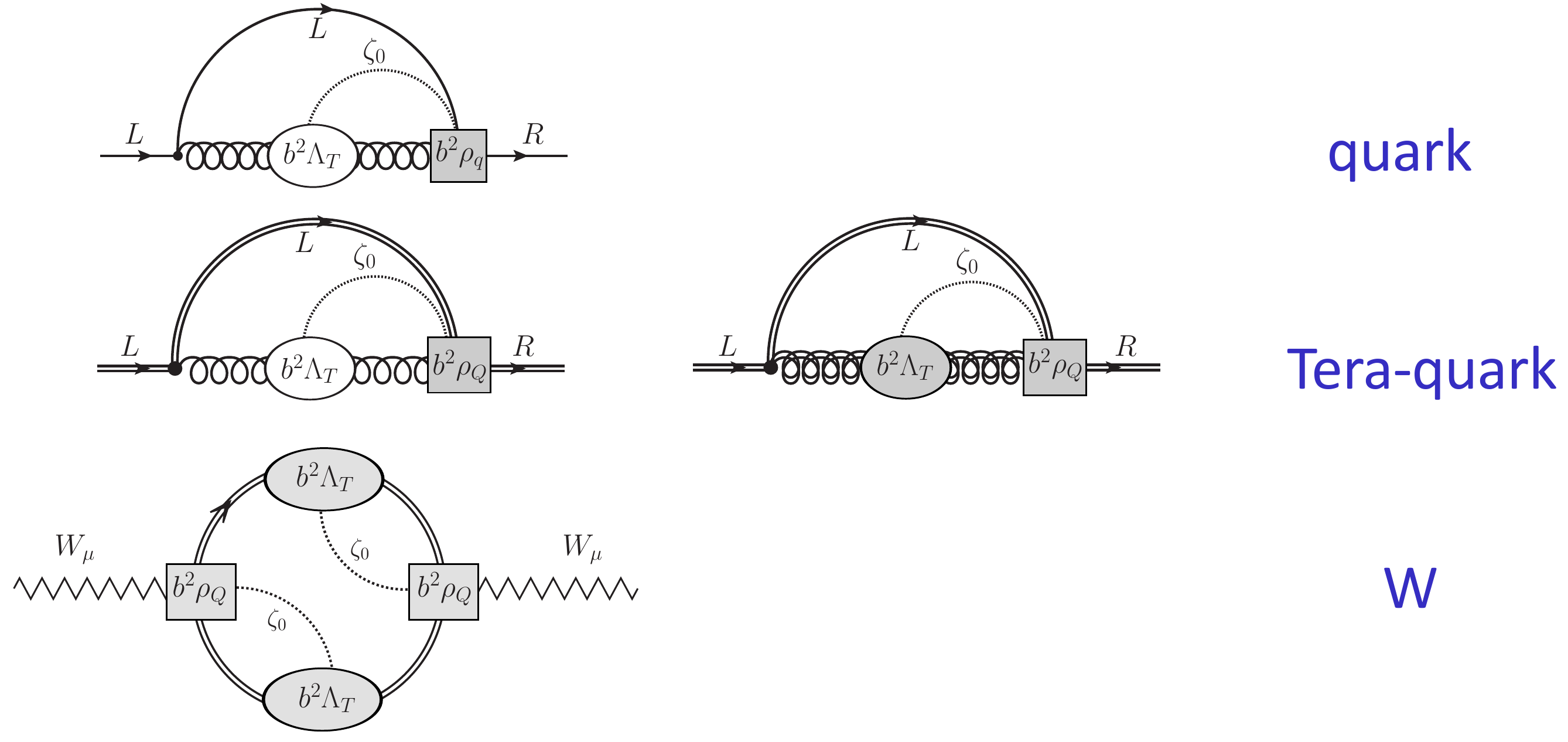}}
\caption{\small{Examples of self-energy diagrams giving NP masses to quarks, Tera-quarks and $W$. The blobs represent the insertion of the NP Symanzik operators~(\ref{OAA}), (\ref{OGG}) and~(\ref{OSS}) for quark and Tera-quark, Tera-quark and $W$, respectively. The rest of the notations is as in figs.~\ref{fig:fig4a} and~\ref{fig:fig5a}.}} 
\label{fig:fig6}
\end{figure}

NP masses emerge from the kind of diagrams shown in fig.~\ref{fig:fig6}. They are all finite owing to the by now well-known UV-IR compensation and all of O($\Lambda_T$) times gauge coupling dependent coefficient functions (see next section). 

\subsection{The critical QEL in the NG phase}
\label{sec:CNGP}

With the same line of arguments we used before to derive eq.~(\ref{GW}), we get for the $d=4$ piece of the QEL of the critical theory in the NG phase the expression 
{\small \beqn
\hspace{-1.4cm}&&{\Gamma}_{4\,cr}^{NG}(q,Q;\Phi;A,G,W)\!=\!
\frac{1}{4}\Big{(}F^A\cdot F^A\!+\! F^G\cdot F^G\!+\!F^W\cdot F^W\Big{)}+\label{GWQCNG}\\ 
\hspace{-1.4cm}&&\quad+\Big{[}\bar q_L \,\,/\!\!\!\! {\cal D}^{WA} q_L+\bar q_R \,\,/\!\!\!\!  {\cal D}^{A}q_R\Big{]}+C_q\Lambda_T\,\Big{(} \bar q_L U q_R+\bar q_R U^\dagger q_L \Big{)}+\nn\\
\hspace{-1.4cm}&&\quad +\Big{[}\bar Q_L\,\,/\!\!\!\! {\cal D}^{WAG} Q_L\!+\!\bar Q_R\,\,/\!\!\!\! {\cal D}^{AG} Q_R\Big{]}+C_Q\Lambda_T\,\Big{(} \bar Q_L U Q_R\!+\!\bar Q_R U^\dagger Q_L \Big{)}+\nn\\
\hspace{-1.4cm}&&\quad+\frac{1}{2}c_w^2\Lambda_T^2{\tr}\big{[}({\cal D}\,^{W}_\mu U)^\dagger {\cal D}^{W}_\mu U\big{]} \nn\, .
\eeqn 
}
From a detailed analysis of the diagrams in fig.~\ref{fig:fig6} one can read-off the parametric dependence of the mass of quarks, Tera-quarks and $W$s to the leading loop order. One finds
\beqn
\hspace{-1.3cm}&&m_q^{NP}\!=\!C_q\, \Lambda_{T}\, , \,\,\,\,\,C_q\!=\!{\mbox{O}}(\alpha_s^2)\label{MQ1}\\
\hspace{-1.3cm}&&m_Q^{NP}\!=\!C_Q \,\Lambda_{T}\, , \,\, \, \,C_Q\!=\!{\mbox{O}}(\alpha_T^2, \ldots)\label{MQ2}\\
\hspace{-1.3cm}&&M_W^{NP}\!=\! C_w\,\Lambda_{T}\, , \,\,\, \,C_w\!=\!\!\sqrt{\alpha_w}\,c_w\, ,\,\, c_w=k_w{\mbox{O}}(\alpha_T,\ldots)\label{MM}\, .
\eeqn

\section{Integrating out Tera-particles and the SM}
\label{sec:ITDF}

We show in this section that, upon integrating out the (heavy) Tera-dof's in the critical model~(\ref{SULLWQ}), the resulting LEEL closely resembles the SM Lagrangian.\ The argument rests on the conjecture that the 125~GeV resonance detected at LHC is a $W^+W^-/ZZ$ composite state bound by Tera-particle exchanges. This state, which we shall denote by $h$, is a singlet under all the symmetries of the theory. Since its mass is $\ll\Lambda_T$,  it must be included in the LEEL, valid for (momenta)$^2\ll\Lambda_T^2$. In these kinematical conditions the most general Lagrangian invariant under the symmetries of the critical model~(\ref{SULLWQ}), in particular under the $\chi_L \times \chi_R$ transformations, and including $h$, takes the form
{\small \beqn
\hspace{-1.3cm}&&{\cal L}_{LEEL}^{NG}=
\frac{1}{4}\Big{(}F^A\!\cdot\! F^A+F^W\!\cdot\! F^W\Big{)}
+\big{(}\bar q_L \,\,/\!\!\!\! {\cal D}^{AW} q_L+ \bar q_R \,\,/\!\!\!\!  {\cal D}^{A} q_R\big{)} 
+ \nn \\
\hspace{-1.3cm}&&+(y_q h + k_q k_v) \,\big{(} \bar q_L U q_R+\bar q_R U^\dagger q_L \big{)} +\frac{1}{2}\partial_\mu h\partial_\mu h +\nn\\
\hspace{-1.3cm}&&+\!\frac{1}{2}(k_v^2\!+\!2k_v k_1 h\!+\!k_2h^2){\tr}\big{[} ( {\cal D}\,^{W}_\mu U)^\dagger {\cal D}^{W}_\mu U\big{]} \!+\!\widetilde {\cal V}(h)\!+\! \ldots \, , \label{VLELA}
\eeqn}
where dots represent $\tilde\chi_L\!\times\!\tilde\chi_R$ violating operators of dimension $d\!>\!4$. The scalar potential $\widetilde{\cal V}(h)$ comprises the cubic and quartic self-interactions of the $h$ field, and includes the $h$ mass term, $m_h^2h^2/2$. The coefficients $k_v, k_1, k_2, y_q, k_q$ and the $\widetilde{\cal V}$-couplings are parameters that need to be fixed by matching onto the underlying (renormalizable and unitary) fundamental critical theory~(\ref{SULLWQ}).

Tree-level matching of ${\cal L}_{LEEL}^{NG}$ (eq.~(\ref{VLELA})) with the QEL $\Gamma^{NG}_{4\, cr}$ (eq.~(\ref{GWQCNG})) requires the identifications
\beqn 
\hspace{-1.2cm}&& k_q k_v = C_q \Lambda_T = m_q^{NP}\, \quad g_w k_v = g_w\, c_w \Lambda_T= M_W^{NP} \, .\label{M2}
\eeqn

Tree-level unitarity implies the constraints
\beq 
y_q=k_q  \, , \qquad k_1 = k_2 = 1 \, .\label{UNIR}
\eeq
These relations are expected to hold up to small loop effects controlled by $g_w$ and $y_q$. Neglecting these corrections, one recognizes that precisely the combination
\beq
\Phi \equiv (k_v+h) U \label{PHIDEF}
\eeq
enters in the $d\leq 4$ part of ${\cal L}_{LEEL}^{NG}$ with the exception of the scalar potential $\widetilde{\cal V}(h)$. Eq.~(\ref{VLELA}) can then be rewritten in the suggestive form
{\small \beqn
\hspace{-1.2cm}&&{\cal L}_{4,LEEL}^{NG}\!=\! 
\frac{1}{4}\Big{(}F^A\!\cdot\! F^A\!+\!F^W\!\cdot\! F^W \Big{)} 
\!+\!\big{(}\bar q_L \,\,/\!\!\!\! {\cal D}^{AW} q_L\!+\! \bar q_R \,\,/\!\!\!\!  {\cal D}^{A} q_R\big{)} \!+ 
 \nn \\
\hspace{-1.2cm}&&+y_q\,\big{(} \bar q_L \Phi q_R\!+\!\bar q_R \Phi^\dagger q_L \big{)}\!+\!\frac{1}{2}{\tr}\big{[} ( {\cal D}\,^{W}_\mu \Phi)^\dagger {\cal D}^{W}_\mu \Phi\big{]} \!+\! \widetilde{\cal V}(h)\, . \label{VLELAMS}
\eeqn}
We see that ${\cal L}_{4,LEEL}^{NG}$ looks very much like the SM Lagrangian with, however, an important difference.\ Since the scalar potential $\widetilde{\cal V}(h)$ is supposed to describe, besides the mass, the self-interactions of the (composite) $h$ field, there is no reason why it should have the same form as the SM Higgs potential. 

\section{Conclusions and Outlook}
\label{sec:CAO}

We have shown that, as an alternative to the Higgs mechanism, elementary particle  masses can be NP-ly generated in a strongly interacting theory where chiral symmetry, broken at the UV cutoff level by irrelevant Wilson-like terms, is recovered at low energy owing to the tuning of certain Lagrangian parameters. Since an exact symmetry prevents the existence of standard mass terms of the kind $m \bar \psi\psi$ and in the critical theory there is no Higgs field around, the mechanism we are advocating allows a neat solution of the mass naturalness problem. In fact, no power divergencies can affect physical masses and masses are ``small'', i.e.\ ${\mbox{O}}(\Lambda_{\rm RGI})$, as the massless theory enjoys an enhanced (chiral) symmetry, consistently with the 't Hooft idea of naturalness~\cite{THOOFT}. 

We have seen that to cope with the magnitude of the $W$ and top mass, a super-interacting sector, gauge invariantly coupled to standard matter, must exists leading to a theory with an RGI scale $\Lambda_{\rm RGI}=\Lambda_T$ of the order of a few TeVs. The simplest model discussed in ref.~\cite{Frezzotti:2014wja} enjoying the NP mass generation mechanism can be straightforwardly extended to incorporate weak interactions and the new Tera-degrees of freedom. In a similar way one could introduce leptons and the hypercharge interaction, an issue that we defer to a forthcoming investigation.

\section*{Acknowledgements} 
\vspace{-.1cm}
We wish to thank R.\ Frezzotti for infinitely many discussions on the issues presented in this talk. Exchanges of ideas with M.\ Garofalo are also acknowledged.

\end{document}